# Modélisation spatiale de la formation des agglomérations dans la zone algéroise

Smicha AIT AMOKHTAR∗, Nadjia EL SAADI∗∗, Yacine BELARBI∗∗∗

∗École Nationale Supérieure Agronomique (Algérie)

Email : s.aitamokhtar@ensa.dz

∗∗École Nationale Supérieure de Statistique et Économie Appliquée (Algérie)

Email : enadjia@gmail.com

∗∗∗ Centre de Recherche en Économie Appliquée pour le Développement (Algérie)

Email : belarbiyacine@yahoo.fr

*Résumé*

*L'objectif de ce travail est d'analyser les dynamiques sous-jacentes à la formation des agglomérations dans la zone algéroise en se référant aux théories de la Nouvelle Économie Géographique (NEG) et plus précisément au travail de Paul Krugman (1991) "increasing returns and economic geography" qui explique les mécanismes de la concentration des activités économiques à travers deux types de forces: des forces centripètes qui encouragent la concentration des activités économiques et des forces centrifuges qui freinent le processus d'agglomération. Ces mécanismes sont traduits en un système d'équations non linéaires dont la résolution par les méthodes analytiques est une tâche très rude, d'où le recours aux méthodes numériques. Nous résolvons numériquement ce système d'équations et présentons des simulations numériques du modèle de Krugman en utilisant des données réelles algériennes. Cette application s'inscrit dans la lignée des travaux empiriques de la NEG et a pour but l'analyse des configurations spatiales émergentes pour une meilleure compréhension du phénomène d'urbanisation dans la zone algéroise.*

**Mots clés**: Agglomération, Nouvelle Économie Géographique, Modèle de Krugman, Simulation numérique, Indice de Gini.

**Abstract**

*The goal of this study is to analyze the dynamics underlying Algiers urban area formation with reference to The New Economic Geography (NEG) theories and more precisely to the paper of Paul Krugman (1991), "Increasing returns and economic geography" which explains the mechanisms of economic activities concentration through two types of forces: centripetal forces enhancing the economic activities concentration and centrifugal forces hindering the agglomeration process. In fact, these mechanisms are translated into a system of nonlinear equations which is very hard to solve analytically. As a consequence, the use of numerical methods is highly advocated. We present some numerical simulations using real Algerian data.*

Classification JEL : I1, R3 ,C6

1. Introduction

Maints modèles ont été construits afin de reproduire la réalité économique dont les questions de concentration spatiale des activités économiques et des populations. Les hypothèses de la théorie néoclassique ne sont pas arrivées à expliquer l'énigme de la dispersion inégale des firmes et des populations. Il a fallu attendre le papier séminal de Krugman (1991) *"increasing returns and economic geography"* qui a donné naissance à la Nouvelle Économie Géographique (NEG) et bouleversé les analyses économiques traditionnelles en donnant à l'espace son ampleur.

Paul Krugman (1991) explique la localisation des activités économiques, en faisant recours au modèle de Dixit & Stiglitz (1977). Ce dernier était à l'origine de l'introduction des imperfections du marché et notamment des rendements croissants dans la modélisation des structures des marchés. En effet, Dixit et Stiglitz (1970) formalisent la conception de la concurrence monopolistique de Chamberlin (1933). L'existence de coûts fixes et la diversité des biens intermédiaires conduisent à une structure industrielle telle que chaque entreprise produit un bien différencié tout en restant en situation de concurrence étant donné l'existence de substituts à ce bien. L'idée de Chamberlin est de supposer que les firmes produisent des biens différenciés pour satisfaire la préférence des consommateurs pour la variété. Le modèle proposé par Dixit & Stiglitz (1977) abandonne l'hypothèse de rendements constants et suggère d'envisager la production industrielle comme sujette à des rendements croissants. C'est le fondement de la NEG. Krugman (1991 ; 1993 ; 1995 ; 1998) et Fujita, Krugman & Venables (1999) mettent en exergue l'effet des coûts de transport et considèrent les agglomérations comme une résultante de forces centripètes qui stimulent la concentration des activités économiques et de forces centrifuges qui incitent à la dispersion de ces activités. Ces forces qui agissent dans des sens opposés sont expliquées essentiellement par les effets suivants (Baldwin *et al.* (2002)) :

– L'effet amont ou le « backward linkage », tel qu'il est dénommé par Krugman, vient du fait que les firmes préfèrent se localiser dans les régions où le marché local est large « the home market » afin de bénéficier des débouchés pour leur production.
– L'effet aval dérive de la modification de la distribution spatiale de la production. L'implantation d'une nouvelle firme permet d'intensifier la concurrence entre les firmes et élargit le bassin d'emploi, ce qui attire les travailleurs à migrer vers cette région. Une autre conséquence de cette implantation est la réduction de l'indice de prix (niveau de vie meilleur). Cet effet, qualifié de « forward linkage » par Krugman, est un effet d'entraînement par l'offre.

Krugman s'est référé à la théorie de l'auto-organisation pour appréhender les mécanismes d'agglomération (Krugman (1992,1998)). L'auto-organisation est définie comme étant un processus dans lequel l'organisation d'un système augmente automatiquement sans être régie par une source extérieure. Les systèmes auto-organisés ont des propriétés émergentes. Pour Krugman, les éléments du système sont des agents économiques rationnels. Ces derniers échangent des biens et des services, chacun poursuivant un objectif de maximisation de son utilité individuelle. L'ensemble de ces agents est soumis à un processus d'auto-renforcement que l'on peut identifier aux économies d'agglomération. La formation des agglomérations dépend de l'intensité relative des forces centripètes et centrifuges: quand les forces d'agglomération dominent, les firmes tendent à se concentrer et dans le cas opposé, les firmes ont tendance à se disperser. Les modèles de la NEG indiquent que le rapport entre les forces centripètes et centrifuges dépend des coûts de transaction entre les régions.

La NEG s'est également investie dans l'économie urbaine. Sa littérature portant sur les villes tente d'expliquer les raisons qui portent les acteurs et agents économiques à se grouper dans l'espace. Elle tente aussi de comprendre les avantages que retirent ces acteurs de l'agglomération spatiale. La recherche des économies d'agglomération a créé l'apparition de grandes villes (Kamal (2010). Ce phénomène prend une ampleur importante dans les pays en développement qu'il peut devenir une source de préoccupation majeure aussi bien pour les économistes que pour les politiques et surtout pour les aménagistes. L'Algérie ne sort pas de cette règle: des distributions inégales de la population; plus importantes sur le littoral. Quant à la capitale, en 2008, elle a abrité 8.7% de la population totale selon le RGHP (2008). Cette distribution déséquilibrée a créé des problèmes d'aménagement du territoire auxquels l'Algérie fait face.

Le but de ce travail est de comprendre les dynamiques d'urbanisation en Algérie en recourant à la NEG et plus précisément au modèle de Krugman (1991), en ayant égard aux hypothèses d'imperfection du marché et d'intégration des coûts de transport. Nous limitons notre application aux communes de la wilaya d'Alger qui sont au nombre de 57 et celles des wilayas frontalières Blida, Boumerdes et Tipaza. Notre choix s'est porté sur cette zone car elle offre l'avantage de fournir certaines statistiques nécessaires à notre étude. Nous proposons de simuler le modèle de Krugman sur ces communes algéroises et d'analyser les structures spatiales émergentes dans cette zone d'étude.

Cet article est réparti comme suit: dans la section suivante, nous présentons le modèle de Krugman (1991) qui est le modèle de base de la NEG ainsi que ses hypothèses et ses caractéristiques principales. La section 3 est consacrée à la simulation du modèle de Krugman (1999) sur la zone algéroise. La première partie de cette section présente les différents scénarios à simuler, les données et les valeurs numériques utilisées pour les paramètres du modèle. La deuxième partie présente les résultats de nos simulations, ces résultats sont discutés dans la dernière partie de la section 3. La section 4 donne une conclusion.

2. Le modèle central de la NEG

   2.1 Caractéristiques principales du modèle de Krugman

Le modèle de Krugman (1991) suppose que les agglomérations émergent de l'interaction des économies d'échelle, du coût de transport et du facteur de mobilité. Ce modèle est basé sur certaines hypothèses: l'économie est dotée de deux secteurs, le secteur industriel (*M*) et le secteur agricole (*A*). Le secteur industriel produit à rendements croissants un continuum de variétés d'un produit horizontalement différencié, au moyen d'un seul facteur de production qui est le travail qualifié. Le secteur agricole produit un bien homogène, avec rendements constants, en utilisant le travail non qualifié comme seul facteur.

Les préférences des consommateurs sont supposées les mêmes pour tous les travailleurs et sont décrites par une fonction d'utilité du type Cobb&Douglas. Chaque consommateur maximise son utilité en consommant une combinaison de deux types de biens :

$$\mathcal{U} = C_M^{\mu} C_A^{1-\mu} \qquad (1)$$

où $C_M$ représente la consommation des biens manufacturiers, $C_A$ est la consommation du bien agricole et µ ($0 < \mu < 1$) est une constante représentant la part des dépenses dans les biens manufacturiers. Par conséquent, $(1 - \mu)$ représente la part des dépenses consacrée aux biens agricoles. Il est supposé que la fonction de consommation des biens industriels $C_M$ est une fonction du type CES[1] telle que :

$$C_M = \left[\sum_{i=1}^{N} c(i)^{(\sigma-1)/\sigma}\right]^{\sigma/(\sigma-1)} \qquad (2)$$

avec $\sigma > 1$.

Dans cette spécification, le paramètre $\sigma$ représente l'élasticité de substitution entre les variétés du bien industriel et N le nombre de variétés. $c(i)$ représente la quantité consommée de la variété i du produit industriel. Par ailleurs, l'espace est supposé composé de n régions où chaque variété est produite dans une seule région. Les coûts de transport sont supposés du type " iceberg ". Cette hypothèse introduite par Samuelson (1954) considère qu'une partie ou un ratio du bien transporté entre deux localisations se perd, au cours du chemin. Donc si $X(i)_{jk}$ est la quantité de la variété *i* exporté de la région *j* vers la région *k*, la quantité de la variété *i* arrivée à la région *k* est :

$$Z(i)_{jk} = e^{-\tau D_{jk}} X(i)_{jk} \qquad (3)$$

où $\tau$ représente le coût de transport[2] et $D_{jk}$ est la distance entre les régions *j* et *k*.

Chaque variété est supposée être produite dans une seule région, et un bien industriel *i* produit dans la région *j* au prix $p(i)_j$ est vendu dans la région *k* au prix :

$$p(i)_{jk} = p(i)_j \, e^{\tau D_{jk}} \qquad (4)$$

Les biens industriels produits dans la même région sont supposés avoir le même prix. Sachant que l'indice des prix varie d'une région à l'autre, l'indice des prix dans la région *k* s'écrit:

$$T_k = \left(\sum_{i=1}^{n} n_i (p(i) e^{\tau D_{ik}})^{1-\sigma}\right)^{1/1-\sigma} \qquad (5)$$

avec $n_i$ le nombre de variétés produites dans la région i et $p(i)$ le prix de chaque variété dans cette région. L'économie est dotée de $L^A$ travailleurs non qualifiés et de $L^M$ travailleurs qualifiés. Les nombres $L^M$ et $L^A$ sont supposés fixes. $\varphi_j$ représente la part des agriculteurs dans la région *j* et $\lambda_j$ est la proportion des ouvriers dans cette région, ces proportions évoluent en fonction du temps. Le secteur agricole produit un seul produit homogène sous des rendements d'échelles constants selon l'équation suivante :

$$L^A_j = q^A_j \quad (6)$$

---

[1] CES: constant elasticity substitution.

[2] La notion de coûts de transport est ici large, elle englobe tous les coûts liés au franchissement d'une distance, aux assurances ( *Gaigné C., F. Goffette Nagot (2008)*).

Le secteur industriel produit les biens manufacturiers sous des économies d'échelle croissantes. En supposant que le seul facteur de production est le travail, la quantité produite est décrite par la technologie suivante :

$$L^M_j = \alpha + \beta\, q(i)_j \quad (7)$$

où $\alpha$ définit le coût fixe en travail. Ce paramètre capte l'effet des économies d'échelle internes, puisque l'expansion de production entraîne une baisse en besoin moyen du facteur travail. $\beta$ définit le coût marginal du travail. $q(i)_j$ est la quantité de la variété $i$ produite dans la région $j$ et $L^M_j$ représente le nombre de travailleurs qualifiés dans la région $j$.

Sous l'hypothèse des économies d'échelle croissantes et la préférence des consommateurs aux variétés, les entreprises sont supposées produire une seule variété. Cette variété est produite dans une seule région. Soit une firme particulière produisant une variété spécifique $i$ dans une région $j$ offrant un taux salarial égal à $W_j$. Du fait que cette dernière entreprise produise une variété unique de biens, cela lui donne un pouvoir de monopole qui sert à maximiser son profit. La condition de maximisation du profit de cette firme en quantité implique que le prix fixé pour la variété $i$ produite dans la région $j$ doit être :

$$p(i)_j = \frac{\sigma}{\sigma - 1} \beta W_j \quad (9)$$

Cette formule indique le fait que si une entreprise réalise un profit positif en produisant une variété d'un bien manufacturé, il est évident que ce secteur motivera d'autres firmes à y investir tout en produisant d'autres variétés. Ainsi, la part du marché de la firme déjà en place commence à baisser: ce phénomène est du à la substitution des variétés.

À long terme, la réalisation des profits positifs incite les firmes à entrer sur le marché alors que la réalisation de pertes incite d'autres à le quitter. Cette libre entrée et sortie sur le marché implique qu'à l'équilibre, le profit est nul et donc la quantité produite à l'équilibre est constante et elle est égale à :

$$q^* = \alpha(\sigma - 1)/\beta \quad (10)$$

2.2 Equilibre à court terme

Deux points essentiels sont induits du modèle de Dixit& Stiglitz (1970). Premièrement, le producteur de chaque variété fait face à une élasticité constante de la demande. Deuxièmement, les firmes réalisent à long terme des profits nuls. Nous induisons que le prix F.O.B.[3] d'un bien manufacturier dans la région $j$ est donné comme suit :

$$P_j = W_j \quad (11)$$

et admettons que les coûts de transport des produits agricoles sont nuls et que le salaire des agriculteurs est le même dans toutes les régions et sera utilisé comme numéraire (on le prend égal à 1) (Krugman (1991)).

*Le revenu dans chaque région*
Soit $\mu$ la proportion d'ouvriers dans l'économie et $(1 - \mu)$ la proportion d'agriculteurs. Le revenu de la région $j$ est donné par :

$$Y_j = (1 - \mu)\varphi_j + \mu\lambda_j W_j \quad (12)$$

---
[3] Free On Bord (FOB): prix de la marchandise sans les coûts de transport et assurances..etc.

*L'indice des prix des produits manufacturiers :* Nous notons que le prix C.i.f [4] d'une unité importée d'une variété de la région $k$ vers la région $j$ est égal à $W_k e^{(\tau D_{jk})}$. Soit $T_j$ l'indice des prix dans la région $j$, alors :

$$T_j = \left[\sum_{k=1}^{n} \lambda_k (W_k e^{\tau D_{jk}})^{1-\sigma}\right]^{1/1-\sigma} \qquad (13)$$

Le salaire nominal à l'équilibre est alors :

$$W_j = \left[\sum_{k=1}^{n} Y_k \left(e^{\tau D_{jk}}\right)^{1-\sigma} T_k^{\sigma-1}\right]^{1//\sigma} \qquad (14)$$

Cette dernière équation est très importante pour nos analyses, puisqu'elle donne le salaire industriel auquel les firmes, dans chaque emplacement, atteignent l'équilibre, étant donné les revenus et indices des prix au sein de chaque région. Nous savons que l'indice des prix est décroissant avec le nombre de variétés (ou de firmes). De ce fait, si le nombre de firmes diminue, ou si la concurrence diminue, alors l'indice de prix augmente et, par conséquent, le salaire également. L'équation (14) détermine le salaire nominal alors que les ouvriers s'intéressent au salaire réel. Ce dernier est défini par:

$$\omega_j = W_j T_j^{-\mu} \qquad (15)$$

### 2.3 Mobilité des facteurs de production et équilibre à long terme

Finalement, nous retournons au facteur de mobilité. Les agriculteurs sont immobiles et les ouvriers sont parfaitement mobiles et ils sont prêts à se déplacer vers les régions où le salaire réel offert est meilleur en se référant au salaire moyen défini par :

$$\bar{\omega} = \sum_{j=1}^{n} \lambda_j \omega_j$$

La fonction de mobilité des firmes est donnée par :

$$\frac{d\lambda_j}{dt}(t) = \rho \lambda_j (\omega_j(t) - \bar{\omega}(t))$$

avec $\rho$ un paramètre reflétant le degré de sensibilité des ouvriers à la migration vers la région offrant le plus grand différentiel de salaires réels.

Le cadre analytique du modèle de Krugman peut être résumé par le système d'équations non linéaires suivant :

$$\begin{cases} \dfrac{d\lambda_j}{dt}(t) = \rho \lambda_j \left(\omega_j(t) - \bar{\omega}(t)\right) \\ \omega_j = W_j T_j^{-\mu} \\ W_j(t) = \left(\sum_k Y_k(t)\left(T_k(t)e^{-\tau D_{jk}}\right)^{\sigma-1}\right)^{1/\sigma} \qquad (I) \\ Y_j(t) = (1-\mu)\varphi_j + \mu \lambda_j W_j \\ T_j(t) = \left[\sum_k \lambda_k (W_k(t)e^{\tau D_{jk}})^{1-\sigma}\right]^{1/1-\sigma} \end{cases}$$

---

[4] Cost Insurance Freight(CIF) : prix de la marchandise comprenant les coûts de transport et assurances..etc.

Cet ensemble d'équations détermine, pour chaque région $j$ ($1 \leq j \leq N$) et à tout instant t (t>0), le niveau du revenu $Y_j(t)$, l'indice des prix $T_j(t)$, le salaire nominal $W_j(t)$ et le salaire réel $\omega_j(t)$. Le système d'équations (I) est complexe pour une résolution analytique d'où le recours aux méthodes numériques (Ait Amokhtar (2012)). En réalisant des simulations numériques, nous explorons les équilibres à long terme du modèle de Krugman.

3. Simulation du modèle de Krugman et application sur la région algéroise

Comprendre les mécanismes d'agglomération est l'objet essentiel du modèle de Krugman (1991), mais ce dernier s'est limité à une économie de deux régions. Le Racetrack economy (Krugman (1992)) illustre la dynamique de la concentration des activités économiques pour un espace composé de n régions. Selon les hypothèses du modèle, les échanges économiques s'établissent sur un cercle. La simulation du racetrack economy montre la possibilité d'avoir plusieurs points d'équilibre avec 12 régions, une structure spatiale avec deux cités ou trois est la plus émergeante. Sur 60 % des cas de simulation, le racetrack economy se termine à se concentrer dans deux régions éloignées de 5, deux concentrations éloignées de 6 pratiquement dans les autres cas. À des intervalles rares, un des résultats mènerait à trois concentrations équidistantes.

Les simulations du modèle de Krugman sur des données réelles ont été initiées par Dirk Stelder (2005). Ce dernier développe ce modèle dans le cas bidimensionnel afin de tester sa robustesse dans la prédiction de la formation des agglomérations sur des structures réelles (Dirk (2005)).

Notre application s'inscrit dans la lignée des travaux de Dirk (2005) et Hakan (2010) avec le but de mener une analyse prospective sur l'armature urbaine de la zone algéroise. La simulation du modèle de Krugman (1991) pour des données réelles représente certaines difficultés dans le sens où il y a manque d'information sur certaines variables du modèle. Choisir des variables proxys est devenu la solution la plus utilisée dans les modèles microéconomiques (Rappaport (1999), Tiebot (1956)). Dans notre cas, nous admettons que la répartition initiale des travailleurs qualifiés dans les différentes communes est proportionnelle à la répartition des entreprises. En outre, nous supposons que le nombre de firmes actives sur le marché algérien est égal au nombre de firmes souscrites au niveau de la société nationale d'électricité et du gaz SONELGAZ voire que cette dernière est la seule compagnie de la distribution d'électricité qui est active en Algérie. Une autre variable pour laquelle nous avons enregistré un manque d'information est le taux salarial par commune, par conséquent, les salaires nominaux initiaux seront assimilés aux valeurs des taux d'urbanisation (Belarbi, 2009). Les données considérées dans notre simulation sur la proportion des agriculteurs dans les différentes communes $\varphi_j$ sont issues de l'office national de statistiques (2008). La matrice des distances intercommunales est calculée à partir des coordonnées polaires (latitude et longitude) de chaque commune.

Trois paramètres sont considérés capitaux dans notre étude, à savoir, la part des biens industriels dans les dépenses de consommation ($\mu$), l'élasticité de substitution entre les variétés ($\sigma$) et le coût de transport ($\tau$). Nous choisissons pour ces paramètres des valeurs qui sont justifiées par des études empiriques antérieures (Billard (2006)). Nous proposons 12 scénarios (voir Tableau 01) afin de couvrir différentes situations, chaque simulation est comparée à une situation initiale qui est celle de l'année de référence 2008 (AIT-AMOKHTAR (2012)). L'ampleur des forces d'agglomération et de dispersion dépend des valeurs prises par les paramètres $\mu, \sigma$ et $\tau$. Nous rappelons ci-dessous les définitions de ces paramètres et essayons de justifier le choix des valeurs que nous leur avons choisies pour les simulations.

3.1 Valeurs des paramètres $\mu, \sigma$ et $\tau$ et scénarios testés
- *La part des biens manufacturés dans les dépenses de consommation ($\mu$)*

La part du revenu dépensée en biens industriels, représentée par la constante $\mu$ dans la formule (1), a une relation positive avec les forces d'agglomération. Les régions périphériques trouvent des difficultés pour attirer le secteur industriel lorsque ce paramètre est fort. La constante $\mu$ représente une force d'agglomération présente dans les modèles d'économie géographique. Ainsi, plus la part des biens manufacturés dans les dépenses est importante, plus le secteur industriel joue un rôle majeur dans le développement économique de la région.

En se rapportant au travail de Billard (2006), maints articles et travaux font varier la valeur de $\mu$ entre 0.3 et 0.5. Krugman (1991) et Andersson et Forslid (2003) prennent comme valeur de base $\mu = 0.3$. Fujita, Krugman et Venables (1999) et Brakman et al. (2001) tiennent une valeur de $\mu$ égale à 0.4. Selon le rapport de la Conférence des Nations Unies sur le Commerce et le Développement (CNUCED, 2008), le poids du secteur industriel dans des économies moins avancées est trop faible, il atteint une valeur moyenne qui ne dépasse pas 0.12. En nous basant sur ces études antérieures, nous décidons de varier le paramètre $\mu$ sur l'ensemble des trois valeurs (0.1, 0.3 et 0.5).

- *L'élasticité de substitution* ($\sigma$)

L'élasticité de substitution $\sigma$ joue un rôle important dans la NEG. D'abord, elle interprète le degré de préférence des consommateurs pour la variété et la possibilité de substituer un bien à un autre pour un même niveau de satisfaction. En outre, le paramètre $\sigma$ montre aussi le degré de concurrence car plus les variétés sont substituables les unes aux autres, plus l'industrie est concurrentielle.

Plusieurs auteurs se réfèrent au travail d'Hummel (1999 ; 2001) pour l'estimation de $\sigma$. Dans le cadre d'un modèle avec un seul secteur de production, Hummel (2001) trouve une élasticité de substitution comprise entre 2 et 5.26. Dans le cadre d'un modèle à deux secteurs de production, Hummels (2001) étudie l'élasticité de substitution pour 62 types de biens. Il trouve des résultats pour 57 d'entre eux avec une valeur moyenne de $\sigma$ égale à 5.6. Hummels propose aussi une élasticité de substitution comprise entre 2 et 5,26 s'il ne considère qu'un seul secteur de production, mais il conclut à la sous-estimation du paramètre. L'analyse fondée sur un modèle à deux secteurs donne une élasticité de substitution moyenne de l'ordre de 5.6. Nous retenons deux valeurs pour $\sigma$ qui représentent les extrêmes de l'intervalle de Hummel : 2 et 5.

- *Le niveau des coûts de transport* ($\tau$)

Les valeurs retenues pour $T = e^{\tau}$ dans l'ouvrage de référence de Fujita, Krugman et Venables (1999) sont comprises entre 1.5 et 2.1. Billard (2006) fixe le coût de transport des échanges entre les pays de l'Union Européenne pour un niveau plus faible (T=1,093), elle justifie son choix par l'existence d'un marché commun entre les Etats. Nous avons choisi deux valeurs du coût de transport : un niveau faible ($\tau = 0.01$) et un niveau important ($\tau = 0.1$).

Nous proposons d'étudier les configurations spatiales possibles de la zone algéroise à long terme; comment le coût de transport peut affecter la répartition de nos ressources humaines; aussi l'effet des économies d'échelles sur la spécialisation des villes. Le choix de cette zone d'étude n'est pas aléatoire; Alger est la capitale économique et politique du pays, et son aire d'influence englobe les wilayas limitrophes tels que Blida, Boumerdes et Tipaza, avec des déplacements pendulaires et des mobilités quotidiennes de type navette. Aussi, la disponibilité des séries statistiques sur la wilaya d'Alger et les wilayas limitrophes a motivé cette application.

*Tableau 01 : Les scénarios simulés*

| Scénarios | Coût de transport faible $\tau = 0.01$ | Scénarios | Coût de transport important $\tau = 0.1$ |
|---|---|---|---|
| Scénario A.1 | $(\mu, \sigma) = (0.3; 2)$ | Scénario B.1 | $(\mu, \sigma) = (0.3; 2)$ |
| Scénario A.2 | $(\mu, \sigma) = (0.3; 5)$ | Scénario B.2 | $(\mu, \sigma) = (0.3; 5)$ |
| Scénario A.3 | $(\mu, \sigma) = (0.5; 5)$ | Scénario B.3 | $(\mu, \sigma) = (0.5; 5)$ |
| Scénario A.4 | $(\mu, \sigma) = (0.5; 2)$ | Scénario B.4 | $(\mu, \sigma) = (0.5; 2)$ |
| Scénario A.5 | $(\mu, \sigma) = (0.1; 2)$ | Scénario B.5 | $(\mu, \sigma) = (0.1; 2)$ |
| Scénario A.6 | $(\mu, \sigma) = (0.1; 5)$ | Scénario B.6 | $(\mu, \sigma) = (0.1; 5)$ |

3.2 Résultats des simulations

Les résultats de nos simulations sont répartis en deux points (3.2.1 et 3.2.2); le premier groupe rassemble les scénarios où le coût de transport a une faible valeur ($\tau = 0.01$). Ces scénarios exposent une situation où l'État adopte une politique publique en infrastructure du transport (Mansouri (2008)). Le deuxième point regroupe les simulations dont les coûts de transport sont importants.

3.2.1 Simulation du modèle de Krugman pour des coûts de transport faibles

Fig.1 présente le résultat de simulation du scénario A.1 où la part de l'industrie est intermédiaire alors que le degré de substitution et le coût de transport sont faibles ($\mu = 0.3$; $\sigma = 2$; $\tau = 0.01$). Nous remarquons pour ce scénario une dispersion des firmes du centre vers les périphéries. À l'équilibre, $\sigma$ mesure les économies d'échelle: un degré de substitution faible traduit une hausse des économies d'échelle et une faible concurrence entre les entreprises, ce qui stimule ces dernières à s'agglomérer. En outre, la baisse du coût de transport motive les travailleurs et les firmes à s'implanter en régions périphériques sans perdre les avantages liés à l'agglomération. Nous avons aussi supposé un niveau d'industrialisation faible, ce qui peut entraîner la dispersion des consommateurs vers les périphériques. Nous sommes donc en présence de forces qui vont dans des sens différents; la résultante de ces forces est illustrée dans Fig 1 où nous constatons une dispersion du secteur industriel vers les périphéries.

*Fig .1: Simulation du scénario A.1*

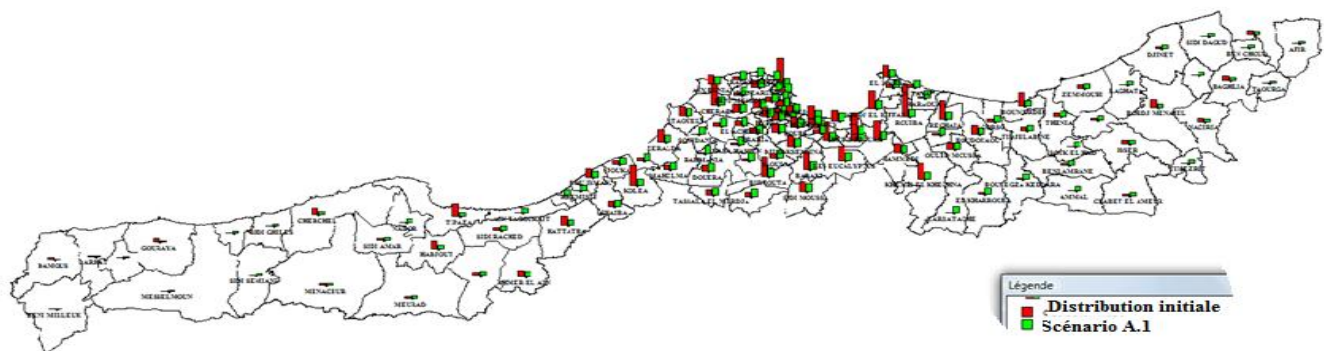

Source: Ait Amokhtar (2012)

Nous avons également simulé le modèle de Krugman pour un taux de substitution important ($\sigma = 5$) (voir Fig. 2). Selon la théorie économique, plus la valeur de $\sigma$ est importante, plus les biens sont substituables, ce qui induit la diminution de la part de marché de chaque firme dans la région où elle s'est installée. Ce phénomène incite les firmes à se délocaliser pour fuir de la concurrence des autres firmes. Cette concurrence résultant aussi de la préférence des ménages pour la variété implique une dilution de la demande quand le nombre de variétés augmente.

*Fig.2: Analyse comparative des scénarios A.1 et A.2*

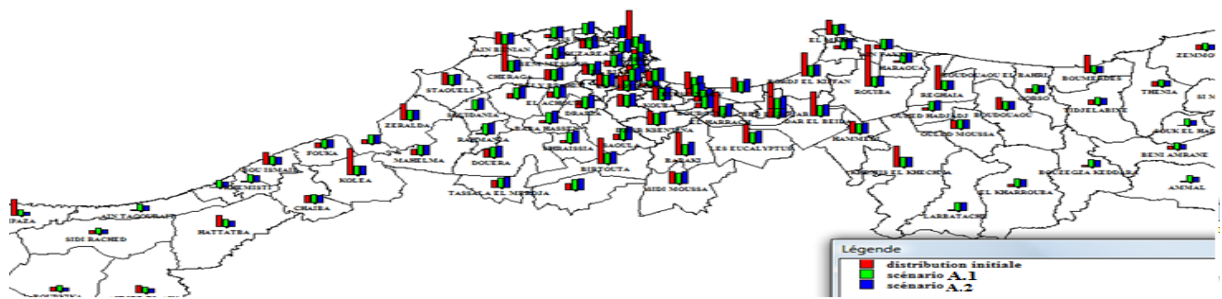

Source: Ait Amokhtar (2012)

D'après les résultats de la simulation du scénario A.2 ($\mu = 0.3; \sigma = 5, \tau = 0.01$), nous observons la même dynamique que le scénario précédent, mais avec la différence que dans le scénario 2, l'intensité de la dispersion des firmes est plus importante. Nous notons ici une disparité importante entre les communes suite à l'augmentation de 2 à 5. La commune de Tipaza enregistre une dispersion des firmes vers les communes qui ont des proportions initiales faibles et cette dispersion s'est accentuée suite à l'augmentation de $\sigma$. Ceci s'explique par le fait que la concurrence due à l'augmentation de $\sigma$ rend plus fort que dans le scénario A.1 le phénomène de dispersion des firmes vers les périphéries. Nous analysons, ci-après, l'effet d'une augmentation des dépenses des produits industriels sur la configuration spatiale des activités économiques en gardant l'hypothèse de la baisse des coûts de transport.

Pour un coût de transport faible, un taux d'industrialisation et un degré de substitution importants (scénario A.3), nous constatons que les firmes ont tendance à s'agglomérer dans un nombre limité de régions (communes), principalement dans la wilaya d'Alger où le marché est grand (Fig.3).

*Fig.3: Simulation du scénario A. 3*

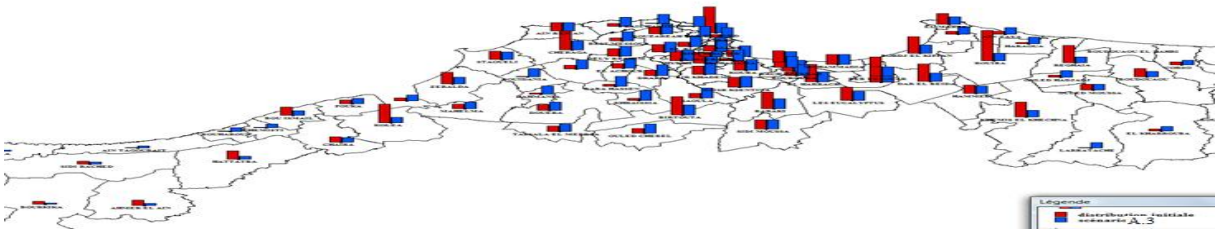

Source : Ait Amokhtar (2012)

Le résultat de ce scénario peut s'expliquer par l'effet de la taille de marché « the home market effect ». Laussel & Paul (2005) donne une brève définition de l'effet de la taille de marché de la manière suivante: « lorsque des biens sont produits avec des rendements d'échelle croissants et que leurs échanges sont soumis à des coûts de transport, les entreprises produisant ces biens ont intérêt à se localiser sur l'aire de marché la plus étendue afin d'économiser les dépenses liées aux transports des biens ».

3.2.2    Simulations du modèle de Krugman pour des coûts de transport importants

Pour un coût de transport important et différentes valeurs du taux de substitution et du taux d'industrialisation, le modèle de Krugman (1991) prévoit une concentration du secteur industriel dans quelques communes. Fig.4 montre significativement l'effet du paramètre coût de transport sur la concentration spatiale du secteur industriel. Les entreprises veulent s'implanter dans la région où se trouve le plus vaste marché afin d'avoir le maximum de débouchés sans supporter des coûts de transport. La taille du marché dépend du nombre de résidents dans cette région et de leurs revenus. Néanmoins, le nombre de résidents est lui-même fonction de la demande de travail formulée par les firmes et donc de la quantité d'emplois disponibles dans la région. Ainsi, la taille du marché dépend du nombre de firmes implantées dans la région. Il faut noter toutefois que le nombre d'entreprises présentes dans la région est lui-même fonction de la taille du marché.

*Fig.4: Simulation du scénario B.1*

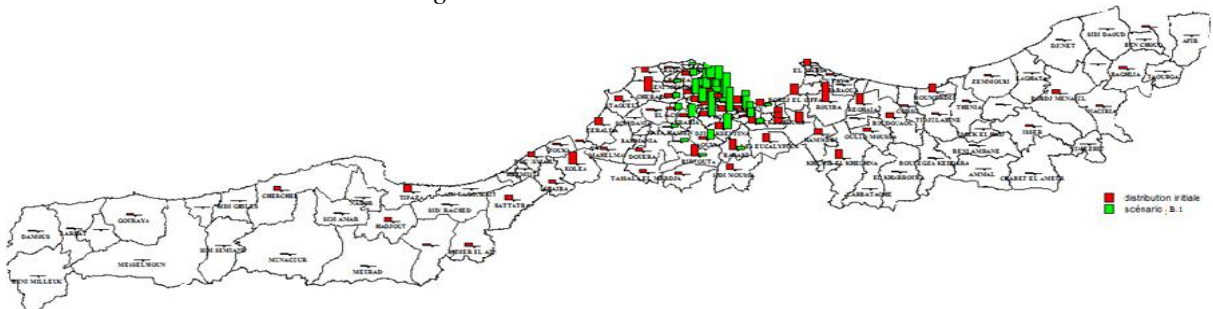

Source: Ait Amokhtar (2012)

L'effet des paramètres $(\mu, \sigma)$ reste faible dans le cas d'un coût de transport bas, par contre il est plus important pour $\tau$ fort. En effet, dans les scénarios reflétant des situations où les coûts de transport sont élevés, la structure est très concentrée et cette concentration s'intensifie lorsque la part de l'industrie augmente et les économies d'échelles deviennent faibles (Ait Amokhtar (2012)).

3.2.3 Indice de Gini

Pour quantifier l'effet de la variation des paramètres $(\mu, \sigma, \tau)$ sur la dynamique de la concentration du secteur industriel, nous recourons au calcul d'un indice de concentration qui est l'indice de Gini. Cet indice donne une mesure de la concentration par rapport à une région de référence qui est la distribution uniforme. L'indice de Gini est calculé comme suit :

$$G = 1 - \sum_{i=1}^{n} \frac{1}{n}[\lambda_i + \lambda_{i+1}]$$

où n représente le nombre de régions et $\lambda_i$ la part de l'emploi du secteur industriel dans la région i après avoir ordonné les régions selon un ordre croissant par rapport aux valeurs des $\lambda_i$. Cet indice varie entre 0 et 1. Il prend la valeur 0 dans une situation d'égalité parfaite où les parts de l'emploi dans toutes les régions seraient égales. À l'autre extrême, il est égal à 1 dans la situation la plus inégalitaire possible. Entre 0 et 1, l'inégalité est d'autant plus forte que l'indice de Gini est élevé.

Fig. 5 résume nos résultats de simulation pour tous les scénarios. L'indice de Gini y a été calculé sur la base des valeurs des $\lambda_j$ associées à l'équilibre spatial. Pour $\tau = 0.01$, nous obtenons des valeurs de G faibles, ce qui indique une faible concentration spatiale de l'activité industrielle au niveau des communes. L'indice de Gini atteint des valeurs proches de 1 pour des coûts de transport forts, ce qui signifie une forte concentration de l'activité industrielle dans un nombre limité de régions.

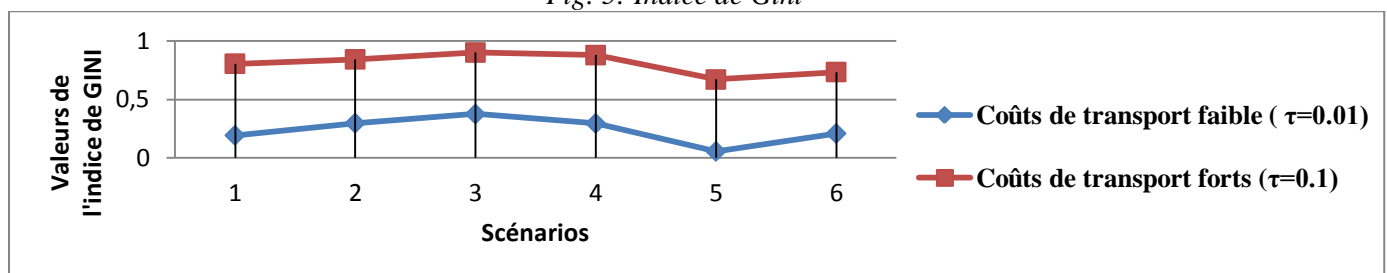

*Fig. 5: Indice de Gini*

Source: Ait Amokhtar (2012)

3.3 Discussion

En nous basant sur les résultats de la sous-section précédente, nous résumons les principaux résultats obtenus.

Notre résultat majeur concerne l'effet de la baisse des coûts de transport sur la disparité spatiale du secteur industriel. Cette baisse a pu corriger les inégalités spatiales existant dans les communes, mais cette convergence entre les régions n'est pas absolue: la commune qui a un pouvoir de marché important à l'état initial reste dominante, mais avec un pouvoir de marché moins fort. En outre, l'importance des coûts de transport exacerbe les disparités spatiales. L'amplification de la concentration en présence de coûts de transport élevés peut être expliquée par le fait que les firmes sont incitées à produire dans les régions qui ont un grand marché afin d'amortir les coûts d'échanges sur une grande part de ventes. Ce résultat sur l'effet des coûts de transport est en accord avec des travaux d'auteurs de la NEG qui se sont investis dans des questions de politique économique où ils étudient l'impact de l'intervention publique sur la localisation des activités entre les régions. Citons par exemple, Jacques-François Thisse et Miren Lafourcade (2008) qui ont analysé l'effet de l'amélioration du réseau de transport sur les inégalités régionales en France. Ils ont déduit qu'une amélioration de la qualité des infrastructures de transport réduira les inégalités régionales.

Un autre résultat important obtenu à partir de nos simulations est l'effet de la distribution initiale des firmes sur leur décision de localisation à long terme. Pour un coût de transport important, les firmes sont poussées à s'installer sur un nombre limité de communes où la taille de ces dernières, en temps initial, est grande. Ces firmes préfèrent également s'installer dans ces communes, car le marché du travail est grand, ce qui leur assure d'une part une main d'œuvre qualifiée. D'autre part, les régions pionnières ont en général développé de meilleurs infrastructures et services offerts aux entreprises, ce qui incite les nouvelles firmes à s'installer. Les consommateurs ont également intérêt à s'installer dans ces communes dites pionnières voire les opportunités d'embauche possibles sur le marché du travail. Cet effet est dénommé dans la littérature « the home market effect » ou l'effet de la taille de marché.

Nous étendons notre discussion au deuxième paramètre clé qui est le taux de substitution ($\sigma$) (ou degré de préférence à la variété). Ce paramètre permet de capter les économies d'échelles et le degré de la différenciation des produits. A l'équilibre, ce taux de substitution a une relation inverse avec les économies d'échelles. L'existence d'un coût fixe de production, auquel est associé un coût marginal supposé constant, favorise l'implantation des activités productives en un lieu unique, à proximité du marché offrant les potentialités marchandes les plus grandes. Les coûts fixes sont ainsi répartis sur un plus grand nombre d'unités produites et vendues et chaque entreprise bénéficie d'un effet taille de marché. Un faible taux de substitution entre les variétés se traduit par une faible concurrence entre les firmes et les consommateurs deviennent indifférents par rapport à leur choix.

Enfin, nous avons retrouvé aussi à partir des scénarios simulés que l'augmentation du $\mu$ a un effet positif sur la concentration spatiale des firmes. Plus la part des dépenses des produits manufacturiers est importante, plus les firmes ont une incitation à s'agglomérer dans un petit nombre de régions afin de bénéficier des gains liés à leur concentration.

4  Conclusion

« Distribution inégale des activités économiques », « coût de transport » et « économies d'échelles » sont considérés comme des concepts essentiels dans la compréhension des mécanismes de formation des agglomérations. La Nouvelle Economie Géographique explique ces mécanismes en se fondant sur le modèle d'équilibre général de Dixit & Stigliz (1977) auquel sont introduits les coûts de transport. Les économistes trouvent impossible d'avoir des agglomérations en se fondant sur les hypothèses classiques. Dans ce travail, nous nous sommes basés sur le modèle de Krugman (1991), modèle fondateur de la Nouvelle Economie Géographique, pour analyser les dynamiques de la formation des agglomérations dans la zone algéroise. La projection du modèle de Krugman sur le cas de la région d'Alger a permis de retrouver un résultat majeur: la baisse des coûts de transport permet de rétrécir les inégalités spatiales existantes entre les régions. A cet effet, il est nécessaire que l'Etat parvienne à intégrer le concept d'agglomération dans sa politique d'attractivité territoriale (par exemple, par une intervention sur le secteur de transport pour assurer une connectivité entre les centres de production et les périphériques, l'Etat réduira les inégalités spatiales dans les différentes communes). L'effet du marché domestique ou « the home market » était aussi visible dans nos résultats de simulations. Quand les coûts de transport sont élevés, les travailleurs qualifiés sont incités à s'installer sur le grand marché où l'opportunité d'embauche est importante et leur préférence à la variété est satisfaite. Quant aux firmes, elles sont disposées à s'implanter dans les grands marchés où la demande est forte et la capacité d'acquisition d'une main d'œuvre spécialisée est meilleure.

Le modèle de Krugman (1991) décrit les relations microéconomiques qui régissent le comportement des agents économiques à travers un modèle d'équilibre général. Ce dernier résume ces relations dans un système d'équations non linéaires. La formalisation mathématique pose certaines hypothèses très restrictives: l'hypothèse de l'uniformité de l'espace et l'homogénéité des comportements des individus, ce qui nous offre des explications partielles par rapport au phénomène d'agglomération. Donc, la question que nous nous posons est la suivante: y a-t-il un substitut à ces méthodes traditionnelles ? Une méthode récente, qui offre une solution aux modélisations traditionnelles et qui peut contribuer à la NEG, est les systèmes multi-agents (Michel, Ferber et Drogoul (2009)); cette approche consiste à créer un monde artificiel composé d'agents en interaction où chaque agent est décrit comme une entité autonome, le comportement des agents est la conséquence de leurs observations, de leurs tendances internes, de leurs représentations et de leurs interactions avec l'environnement et les autres agents (communications, stimuli, action directe, etc..). Nous pensons, dans un travail futur, à une extension du modèle de Krugman (1991) en utilisant une approche multi-agents qui permettrait l'introduction de l'hétérogénéité dans l'espace et dans les comportements des agents économiques et leurs interactions.

Références bibliographiques

*Ait Amokhtar Smicha (2012), « Modélisation spatiale de la formation des agglomérations. Application à des villes algériennes », mémoire de magister en statistique appliquée, ENSSEA.*
*Andersson F et Forslid R (2003), "Tax Competition and Economic Geography", Journal of public Economic Theory, Vol. 5(2), pp. 279-303.*
*Baldwin Richard, Rikard Forslid, et Philippe Martin, Gianmarco Ottaviano et Frédéric Robert (2002), "Economic Geography and Public Policy", Princeton University Press.*
*Belarbi Yacine (2009), « Convergence régionale de l'emploi et dépendances spatiales : Le cas de l'Algérie. Approche par l'économétrie Spatiale », Thèse de Doctorat, Institut National de la Planification et de la Statistique d'Alger & Université JEAN MONNET de SAINT-ETIENNE.*
*Billard Catherine (2006), « Dépenses publiques, localisation des capitaux et concurrence fiscale: une modélisation et économie géographique », Thèse de Doctorat, Economies and Finances, Université Panthéon Sorbonne - Paris I.*
*Brakeman Steven, Garretsen Harry et Van Marrewijk Charles (2009), "The New Introduction to Geographical Economics", Cambridge University, Press Economics.*

*Brian D. Hahn et Daniel T. Valentine (2007), "Essential MATLAB for Engineers and Scientists", Third edition, Elsevier Ltd.*

*Chamberlin Edward (1933), "The theory of Monopolistic Competition", The economic journal, 43 (172), pp661-666.*

*Coissard Steven (2007), « Perspectives de la nouvelle économie géographique de Paul Krugman : Apports et limites », Revue d'Économie Régionale & Urbaine, No 1 - pp. 111-125.*

*Combes Pierre Philippe, Mayer Thierry et Thisse Jacques François (2008), « Economic geography: the integration of regions and nations », Princeton University Press.*

*Darrigues Fabrice et Catt Jean-Marc Montaud (2003), « Les expériences d'intégration latino-américaines à la lumière de la Nouvelle Economie Géographique », Séminaire EMMA-RINOS, Analyse comparatiste des processus d'intégration régionale Nord-Sud, Paris 26-27 Mai 2003.*

*Dixit, Avinash K.; Stiglitz E. Joseph (1977), "Monopolistic Competition and Optimum Product Diversity", The American Economic Review, Vol. 67, No. 3.pp 297-308.*

*Fujita Masahisa, Krugman Paul, Venables Anthony J. (1999), "The spatial economy: Cities, Regions and International Trade", Cambridge, Massachussets: MIT Press.*

*Gaigné C., F. Goffette Nagot (2008), « Localisation rurale des activités industrielles. Que nous enseigne l'économie géographique?», Revue d'Etudes en Agriculture et Environnement 87 pp 101-130.*

*Hakan Andic (2010), « L'impact des politiques de transport sur la concentration spatiale des activités », PhD Thesis, université LAVAL, Québec.*

*Hummels David ( 1999), "Toward a geography of trade costs", GTAP Working Paper No. 17, Purdue University.*

*Hummels David ( 2001), "Toward a geography of trade costs", Mimeo, Purdue University.*

*Kamal Abdelhak (2010), « Industrialisation et concentration urbaine », Thèse de Doctorat, Université du Sud Toulon-Var : Faculté de Sciences Economiques et de Gestion.*

*Krugman Paul (1991), "Increasing returns and economic geography", Journal of Political Economy, 99 (3) pp 483-499.*

*Krugman Paul(1992), "A dynamic spatial model", Working Paper.*

*Krugman Paul (1993), "On the number and location of cities", European Economic Review 37 pp 293-298.*

*Krugman Paul (1995), "Development, Geography and Economic Theory", Cambridge , Massachussets: MIT Press.*

*Krugman Paul (1998), « L'économie auto-organisatrice », De Boeck Université, Bruxelles.*

*Lafourcade Miren et Thisse Jacques-François (2008), "New economic geography: A guide to transport analysis", PSE Working Papers halshs-00586878, HAL*

*Laussel Didier et Paul Thierry (2005), « L'effet taille de marché: un réexamen du modèle de Helpman-Krugman et de quelques extensions », Revue d'économie politique 115 (5).*

*Mansouri Yassine (2008), « La localisation des activités productives: les tensions entre les forces centrifuges et centripètes», Thèse de Doctorat « es sciences économiques », Université du Sud Toulon Var.*

*Maurice Catin, Cuenca Christine, Kamal Abdelhak (2008), « L'évolution de la structure et de la primatie urbaine au Maroc », Région et Développement n° 27.*

*Michel Fabien, Ferber Jacques et Drogoul Alexis (2009), « Multi-Agent Systems and Simulation: a Survey From the Agents Community's Perspective », Multi-Agent Systems: Simulation and Applications (5) pp 3—52.*

*Mossay Pascal et Picard Pierre (2009), "On Spatial Equilibria in a Social Interaction Model », core discussion paper.*

*Rappaport Jordan (1999), "How does labor mobility affect income convergence?", Research Working Paper 99-12, Federal Reserve Bank of Kansas City.*

*Samuelson, Paul A (1954), " 'The Transfer Problem and Transport Costs, II: Analysis of Effects of Trade Impediments", The Economic Journal 64 (254 ) pp 264–289.*

*Stelder Dirk (2005), « Regions and Cities: Five Essays on Interregional and Spatial Agglomeration Modeling », Thèse de Doctorat, Université de Groningue.*

*Teixeira Fernandes (2002), « Transport policies in light of the new economic geography: the Portuguese experience », CORE, Université Catholique de Louvain and Faculdade de Economia do Port.*

*Tiebout Charles M, "A pure theory of local expenditures", Journal of Political Economy 64 (5), pp 416-424.*